\documentclass[pre,aps,amsmath,amssymb,reprint,floatfix,superscriptaddress,longbibliography]{revtex4-1}

\pdfoutput=1
\usepackage{xcolor}
\usepackage{siunitx}
\usepackage{graphicx}
\usepackage[colorlinks=true,citecolor=blue,urlcolor=blue]{hyperref}
\usepackage{etoolbox} 
\usepackage{bm} 
\usepackage[normalem]{ulem}

\newcommand{\thin}{\mskip0.75mu}

\def\cal#1{\mathcal{#1}}
\def\eqq#1{Eq.~(\ref{#1})}

\def\eq#1{(\ref{#1})}

\def\f#1{Fig.~\ref{#1}}

\def\c#1{~\cite{#1}}

\def\cc#1{~Ref.~\cite{#1}}

\def\av#1{\langle #1 \rangle}

\def\beq{\begin{equation}}
\def\eeq{\end{equation}}
\def\bea{\begin{eqnarray}}
\def\eea{\end{eqnarray}}

\def\x{{\bm x}}
\def\tt{{\bm \theta}}

\def\kB{k_{\rm B}}
\def\tf{t_{\rm f}}
\def\kt{\kB T}

\begin{document}

\title{Training thermodynamic computers by gradient descent}

\author{Stephen Whitelam}
\email{swhitelam@lbl.gov}

\affiliation{Molecular Foundry, Lawrence Berkeley National Laboratory, 1 Cyclotron Road, Berkeley, CA 94720, USA}

\begin{abstract}
We show how to adjust the parameters of a thermodynamic computer by gradient descent in order to perform a desired computation at a specified observation time. Within a digital simulation of a thermodynamic computer, training proceeds by maximizing the probability with which the computer would generate an idealized dynamical trajectory. The idealized trajectory is designed to reproduce the activations of a neural network trained to perform the desired computation. This teacher-student scheme results in a thermodynamic computer whose finite-time dynamics enacts a computation analogous to that of the neural network. The parameters identified in this way can be implemented in the hardware realization of the thermodynamic computer, which will perform the desired computation automatically, driven by thermal noise. We demonstrate the method on a standard image-classification task, and estimate the thermodynamic advantage -- the ratio of energy costs of the digital and thermodynamic implementations -- to exceed seven orders of magnitude. Our results establish gradient descent as a viable training method for thermodynamic computing, enabling application of the core methodology of machine learning to this emerging field.
\end{abstract}
 
\maketitle

\section{Introduction}

Thermodynamic computing offers a potential route to energy-efficient computation. Unlike digital or quantum computing, which must at considerable energetic cost overpower or suppress thermal noise, thermodynamic computing is designed to use thermal noise as a source of energy. Physical devices whose states evolve under  Langevin dynamics can be engineered to perform computations as they relax toward thermal equilibrium. Because these computations are carried out by the natural dynamics of the system, such devices can in principle operate with very low energy overhead, approaching fundamental thermodynamic limits\c{conte2019thermodynamic,hylton2020thermodynamic,wimsatt2021harnessing,aifer2024thermodynamic,melanson2025thermodynamic,aifer2025solving}.

A key challenge for thermodynamic computing is to identify algorithms that make efficient use of thermodynamic hardware and that reproduce the algebraic and machine-learning operations done digitally. Recent work has shown that thermodynamic computers can solve linear algebra problems, such as matrix inversion, in thermodynamic equilibrium\c{aifer2024thermodynamic,melanson2025thermodynamic}. The advantage of equilibrium operation is that the computer's degrees of freedom obey the Boltzmann distribution, which depends in a precise way on the computer's potential energy. By choosing this potential energy appropriately, therefore, we can specify the desired computation. In this mode of operation the device functions as a {\em Boltzmann computer}, sampling the Boltzmann distribution~\footnote{Used in this way, a thermodynamic computer resembles a continuous-spin analog of a Boltzmann machine, a statistical mechanical model that represents probability distributions over binary variables\c{hinton2017boltzmann,salakhutdinov2009deep}.}. However, one drawback of equilibrium operation is the requirement to attain equilibrium: physical systems can take a long time to equilibrate\c{cugliandolo2002dynamics,cipelletti2005slow}, and the equilibration time of thermodynamic computer programs can vary by orders of magnitude as the computer's parameters are changed\c{whitelam2024increasing}.

One response to slow equilibration is to devise ways of increasing the basic clock speed of a thermodynamic computer, such as by rescaling its potential energy and injecting noise in suitable proportion\c{whitelam2024increasing}. Another is to design a thermodynamic computer to operate out of equilibrium.  In recent work we showed that a thermodynamic computer can be clocked, performing a desired computation at a specified observation time, regardless of whether equilibrium has been attained\c{whitelam2024thermodynamic,whitelam2025generative}. In this mode of operation the device functions as a {\em Langevin computer}, performing computations via the properties of finite-time Langevin trajectories.

\begin{figure*} 
   \centering
   \includegraphics[width=\linewidth]{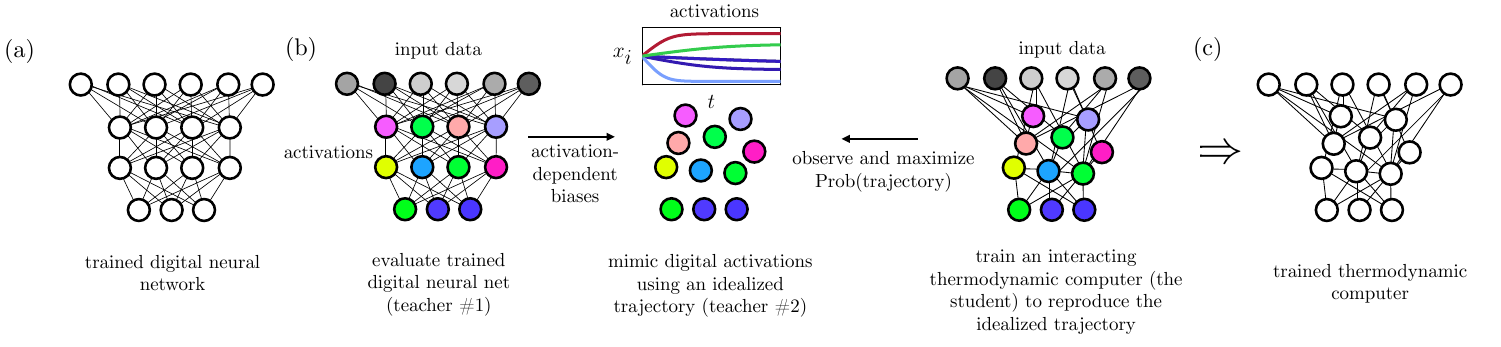} 
   \caption{Training a thermodynamic computer by gradient descent. (a) A trained neural network is used as a reference model. (b) Left: Input data are passed through the neural network, producing node activations $A_i$ (colored circles, ``teacher \#1''). Center: These activations are the targets for an idealized trajectory $\{x_i^{(0)}(t)\}$ (``teacher \#2''), such that the values $x_i^{(0)}(\tf)$ observed at time $\tf$ resemble the $A_i$.  Right: An interacting thermodynamic computer (the ``student'') is trained to maximize the probability of generating this trajectory, using the gradient-descent equations \eq{grad1} and \eq{grad2}. (c) Repeating this process for many input data yields a trained thermodynamic computer, which mimics the behavior of the neural network.}
   \label{fig1}
\end{figure*}

In\cc{whitelam2024thermodynamic} we used a genetic algorithm to identify the parameters of a nonlinear thermodynamic computer capable of performing standard machine-learning computations at finite observation times. While effective, genetic algorithms are slower than gradient descent, the standard training method of machine learning\c{lecun2015deep,schmidhuber2015deep}. The contribution of the present paper is to describe how to train a thermodynamic computer by gradient descent, in order to do calculations on a clock. In doing so we enable application of the core training method of machine learning to thermodynamic computing, and confirm that thermodynamic computers can be operated on a clock, as conventional computers are. Training is done on a digital computer; the resulting parameters can be implemented in a hardware realization of the thermodynamic computer, which will perform the desired computation automatically, driven by thermal noise.

Our training method uses a teacher-student setup within a digital simulation of a thermodynamic computer. By taking derivatives of the Onsager-Machlup functional, the path weight for Langevin trajectories, we maximize the likelihood that the thermodynamic computer generates trajectories that reproduce the activations of a reference model, a trained neural network. In this way we design the dynamics of the thermodynamic computer to perform calculations analogous to those performed by the neural network. We demonstrate the method by training a thermodynamic computer to perform image recognition at a specified observation time.

The authors of\cc{lipka2024thermodynamic} recently introduced a teacher-student scheme for thermodynamic computing based on quantum thermal machines coupled to reservoirs at different temperatures. Their quantum units implement Boolean logic gates by relaxing to non-equilibrium steady states. By contrast, our framework uses a network of classical, nonlinear units in contact with a single thermal bath. The parameters of this network are trained by gradient descent so that its finite-time, clocked dynamics reproduces the activations of a nonlinear neural network. This approach enables classical, programmable thermodynamic computing capable of tackling high-dimensional tasks such as image recognition, extending the scope of thermodynamic computing beyond Boolean logic (and linear algebra) to scalable machine learning.

\section{Training a thermodynamic computer by gradient descent} 

Our model of a thermodynamic computer consists of $N$ classical, real-valued fluctuating degrees of freedom $\x=\{x_i\}$. These degrees of freedom can represent different physical observables in thermodynamic computers realized in hardware, including voltage states in electrical circuits\c{melanson2025thermodynamic}, oscillator positions in a mechanical device\c{dago2021information}, or phases in a Josephson junction\c{ray2023gigahertz,pratt2025controlled}. The computer's units $x_i$ evolve in time according to the overdamped Langevin dynamics
\beq
\label{lang1}
\dot{x}_i= -\mu \thin \partial_i V_\tt(\x)  + \sqrt{2 \mu \thin \kt} \, \eta_i(t).
\eeq 
Here $\mu$ is the mobility parameter, which sets the basic time constant of the computer. For the thermodynamic computers of Refs.\c{aifer2024thermodynamic,melanson2025thermodynamic}, $\mu^{-1}$ is of order a microsecond. For damped oscillators made from mechanical elements\c{dago2021information} or Josephson junctions\c{ray2023gigahertz,pratt2025controlled}, $\mu^{-1}$ is of order a millisecond or a nanosecond, respectively.

 The first term on the right-hand side of \eqq{lang1} is the force arising from the computer's potential energy $V_\tt(\x)$, given a set of parameters (couplings and biases) $\tt=(\{J_{ij}\},\{b_i\})$; note that $\partial_i \equiv \partial/\partial x_i$. These are the parameters we wish to adjust in order to have the computer perform a desired calculation. The second term on the right-hand side of \eqq{lang1} models thermal fluctuations: $\kt$ is the thermal energy scale, and the uncorrelated Gaussian white noise terms satisfy $\av{\eta_i(t)}=0$ and $\av{\eta_i(t) \eta_j(t')} = \delta_{ij} \delta(t-t')$. 

We take the potential energy $V_\tt(\x)$ of the computer to be
\beq
\label{pot}
V_\tt(\x) = \sum_{i=1}^N  \left(J_2 x_i^2+J_4 x_i^4\right)+\sum_{i=1}^N b_i x_i +\sum_{(ij)} J_{ij} x_i x_j.
\eeq
The first sum in \eqq{pot}, which runs over the $N$ units of the computer, sets the response of its individual degrees of freedom. For $J_2>0, J_4=0$ we have a linear model similar to that of\cc{aifer2024thermodynamic}, whose activations in equilibrium are linear functions of their inputs. For $J_2,J_4 >0$~\footnote{Note that the choice $J_2<0, J_4>0$ creates bistable units, analogous to the p-spins in the field of probabilistic computing\c{kaiser2021probabilistic}. The choice $J_2,J_4>0$ creates units with one stable state, analogous to the s-units of thermodynamic computing\c{melanson2025thermodynamic}.} we have a nonlinear model that is the thermodynamic analog of a neural network\c{whitelam2024thermodynamic}. This is the case we focus on in this paper. In general, it is not necessary for the computer's potential to be exactly quartic: we can have a nonlinear model with any non-quadratic potential that is thermodynamically stable (bounded from below). We can consider \eqq{pot} to represent a low-order Maclaurin expansion, in powers of $x_i$, of the potential of such a device. 

 The remaining terms in \eqq{pot} contain the adjustable parameters of the computer. The parameters $b_i$ are input signals or biases applied to each unit. The parameters $J_{ij}$ are pairwise couplings between units, and are motivated by the bilinear interactions of the thermodynamic computers of Refs.\c{aifer2024thermodynamic,melanson2025thermodynamic}. As a consequence, the information flow between units $i$ and $j$ is bidirectional -- units $i$ and $j$ feel the forces $-J_{ij} x_j$ and $-J_{ij} x_i$, respectively -- unlike in a feedforward neural network, whose connections are unidirectional. 
 
Our goal is to use digital simulation to train a nonlinear thermodynamic computer, the thermodynamic analog of a neural network, to mimic the behavior of a neural network trained to perform a desired computation. The parameters identified in this way can then be implemented in the hardware realization of the thermodynamic computer, which can perform the desired computation automatically, driven by thermal noise. 

Our proposed training scheme is shown in \f{fig1}. As sketched in panel (a), we start with a neural network trained for a desired task. The neural network has $N$ nodes, the same number of degrees of freedom as our thermodynamic computer, but with potentially different connectivity. We take the trained neural network (``teacher \# 1''), evaluate it for a particular set of input data, and record the $N$ activations $A_i$ of its degrees of freedom. This step is shown on the left of panel (b).

Next, we construct by hand an idealized dynamical trajectory $\{x_i(t)\}$ of the thermodynamic computer. This step is shown in the center of panel (b). In this idealized trajectory the $N$ activations $x_i$ of the thermodynamic computer evolve with time toward values equal to (or similar to) the activations $A_i$ of the trained digital computer. There are different ways to construct such a trajectory. A natural way to do so is to use a noninteracting thermodynamic computer, setting $J^{(0)}_{ij} =0$ and taking the hidden and output biases $b_i^{(0)}$ proportional to the corresponding activations $A_i$ of the trained neural network. The idealized trajectory can even be deterministic (e.g. we can set the temperature of the noninteracting reference computer to zero, $T^{(0)}=0$): as long as we calculate the probability with which a noisy, finite-temperature thermodynamic computer generates the idealized trajectory, the latter can be constructed for convenience. 

Ultimately, our goal is to have the trained thermodynamic computer reproduce the neural-network outputs, but we also require the computer to learn the net's hidden activations. The neural network relies on its hidden units to compute its outputs, and so we reason that reproducing its hidden-unit behavior will allow the thermodynamic computer to acquire a computational capacity similar to that of the neural network.

This idealized trajectory (``teacher \#2'') is observed by an interacting thermodynamic computer (or ``student''), the device we wish to train. The student computer, which operates at finite temperature $T$ and has connections $J_{ij}$ and biases $b_i$, sees the same input data that produced the activations $A_i$ of the trained neural network. As the teacher trajectory runs, the parameters of the student thermodynamic computer are adjusted by gradient descent in order to increase the likelihood with which it {\em would} generate the idealized trajectory. This step, explained in detail below, is shown on the right of panel (b) of \f{fig1}. 

The training step shown in panel (b) is repeated for many input examples. The result is a trained thermodynamic computer, illustrated in panel (c), whose finite-time dynamics realizes a computation analogous to that carried out by the trained neural network. The computer's trained parameters $\tt$ could then be realized in hardware, encoding the forces that drive the computer to perform the same computation as the neural network in panel (a). When presented with input data, the hardware thermodynamic computer will evolve automatically, driven by thermal noise.

\subsection{Details of the gradient-descent algorithm}

 The gradient descent algorithm sketched in \f{fig1}(b) is designed to adjust the parameters $\tt=(\{J_{ij}\},\{b_i\})$ of an interacting thermodynamic computer in order to maximize the likelihood with which it would generate an idealized trajectory, $\omega = \{\x(t_k)\}_{k=0}^{K}$, observed at discrete times $t_k = k \Delta t$; note that $K = \left\lfloor \tf/\Delta t \right\rfloor$. The idealized trajectory is designed to make the activations $x_i(\tf)$ of the thermodynamic computer at some observation time $\tf$ equal to (or similar to) the activations $A_i$ of a trained neural network (left). Starting with the parameters of the trained computer set to zero, $\tt = {\bm 0}$, we proceed as follows.
\begin{figure*} 
   \centering
 \includegraphics[width=\linewidth]{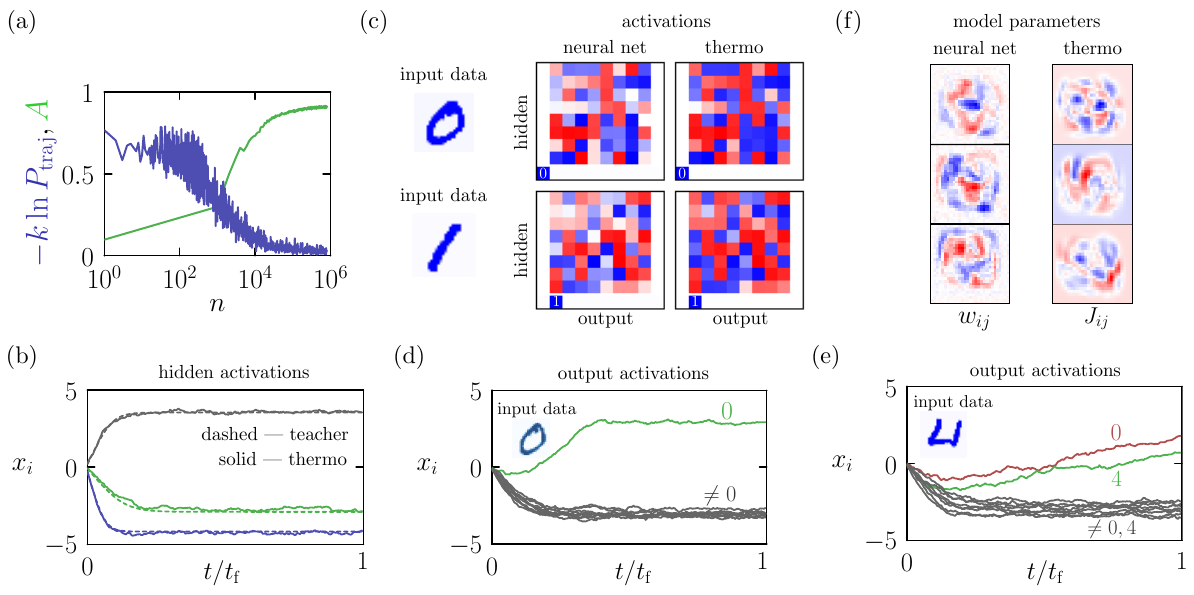} 
   \caption{Training a thermodynamic computer to classify MNIST images. (a) The loss function, the negative log probability that the student thermodynamic computer generated the idealized training trajectory, as a function of the number $n$ of training trajectories, for a 10-class image classifier ($k=10^{-3}$). We also show the validation-set accuracy $A$ of the thermodynamic computer trained using $n$ teacher trajectories. (b) Three activations of the trained computer as a function of time (solid lines, activations distinguished by the three colors), compared with their time dependence in the idealized teacher trajectory (dashed lines). (c) Hidden and output activations given two input digits, for the neural network (left) and the trained thermodynamic computer observed at time $\tf$ (right). The latter values are averaged over 5 trajectories.  (d,e) Activations as a function of time for the thermodynamic computer when connected to the displayed digit. The example in (d) is correctly classified, while the example in (e) is not. (f) Trained neural network weights $w_{ij}$ and thermodynamic computer couplings $J_{ij}$ connecting the input data to 3 of the 64 hidden nodes.}
   \label{fig2}
\end{figure*}

On a digital computer we integrate \eqq{lang1} according to a discretized update equation, such as the first-order Euler scheme
\beq
\label{lang2}
x_i(t+\Delta t)= x_i(t)-\mu \thin \partial_i V_\tt({\bm x})\Delta t  + \sqrt{2 \mu \thin \kt \Delta t} \, \eta_i.
\eeq 
Here $\Delta t$ is the integration timestep, and $\eta_i \sim {\cal N}(0,1)$ is a Gaussian random variable with zero mean and unit variance. We define $\Delta x_i(t)  \equiv x_i(t+\Delta t)-x_i(t)$. Under this integration scheme, the probability of generating a step $\x \to \x + \Delta \x$ of the idealized trajectory is that of drawing the $N$ noise values $\eta_i$, which is
\beq
\label{step}
P^{\rm step}_\tt(\Delta \x)=(2 \pi)^{-N/2} \prod_{i=1}^N \exp(-\eta_i^2/2),
\eeq
where the $\eta_i$ are specified by \eqq{lang2}. Solving \eqq{lang2} for $\eta_i$ and inserting the result into \eq{step} gives the probability that a computer with parameters $\tt$ would have generated the step $\x \to \x +\Delta \x$. The negative logarithm of this expression is
\beq
\label{logp}
-\ln P^{\rm step}_\tt(\Delta \x) =  \sum_{i=1}^N \frac{\left(\Delta x_i+\mu \thin \partial_i V_\tt(\x) \Delta t \right)^2}{4 \mu \thin \kt \Delta t},
\eeq
up to an unimportant constant term. \eqq{logp} is a discrete version of the Onsager-Machlup action associated with the Langevin equation\c{risken1996fokker,cugliandolo2017rules}.

The negative log-probability $-\ln P_{\tt}(\omega)$ with which the computer would generate the entire idealized trajectory $\omega = \{\x(t_{k})\}_{k=0}^{K}$ is then a sum of terms \eq{logp} over each step of the trajectory, $-\ln P_{\tt}(\omega) = -\sum_{k=0}^{K-1} \ln P^{\rm step}_\tt(\Delta \x(t_k))$. This is our loss function. To increase the likelihood with which the thermodynamic computer would generate that trajectory, we update each parameter of the computer by the gradient of the loss with respect to itself, giving
\bea
\label{grad1}
J_{ij} &\to& J_{ij} +\alpha \sum_{k=0}^{K-1} \frac{\partial} {\partial J_{ij}} \ln P^{\rm step}_\tt(\Delta \x(t_k)), \\
\label{grad2}
b_{i} &\to& b_{i} +\alpha \sum_{k=0}^{K-1} \frac{\partial}{\partial b_{i}} \ln P^{\rm step}_\tt(\Delta \x(t_k)) ,
\eea
where $\alpha$ is a learning rate. The gradient terms in \eqq{grad1} and \eqq{grad2} can be calculated analytically, using \eqq{pot}, and are
\bea
\label{grad3}
-\frac{\partial} {\partial J_{ij}} \ln P^{\rm step}_\tt(\Delta \x)&=&   \frac{ \Delta x_i+\mu \thin \partial_i V_\tt(\x) \Delta t}{2 \kt } x_j \\
&+& \frac{ \Delta x_j+\mu \thin \partial_j V_\tt(\x) \Delta t}{2 \kt } x_i \nonumber,
\eea
and
\beq
\label{grad4}
-\frac{\partial} {\partial b_i}\ln P^{\rm step}_\tt(\Delta \x) =   \frac{ \Delta x_i+\mu \thin \partial_i V_\tt(\x) \Delta t}{2 \kt },
\eeq
where
\beq
\label{grad5}
\partial_i V_\tt(\x)=2 J_2 x_i+4 J_4 x_i^3+b_i+\sum_{j \in \mathcal{N}(i)} J_{ij} x_j.
\eeq
Here $\mathcal{N}(i)$ denotes the set of units connected to unit $i$. All quantities $x_i$ in Eqs.~\eq{grad3}--\eq{grad5} are evaluated at time $t_k$, appropriate for each term in Eqs.~\eq{grad1} and \eq{grad2}.

The thermodynamic computer that we are training to reproduce the idealized trajectory sees the same input data as the neural network whose activations we are seeking to copy. This can be accomplished by viewing the input data as a set of frozen variables $x_i$, and coupling them to the degrees of freedom $x_j$ of the thermodynamic computer via the trainable interactions $J_{ij} x_i x_j$ (in this respect the input data function as a set of biases). We repeat the process for a range of input data, and record the resulting parameters $\tt$ of the thermodynamic computer. 

\section{Training a thermodynamic computer to classify images}

To illustrate the method, we applied it to MNIST digit classification\c{mnist_leaderboard}. To create the teacher model, we trained a fully connected neural network with two hidden layers of 32 nodes each, hyperbolic tangent activations, and 10 output nodes (one per digit class). Using stochastic gradient descent (learning rate $0.2$, weight decay $10^{-4}$, batch size $256$) for 300 epochs, the network achieved a test-set accuracy of 97.3\%.

The thermodynamic computer consists of 64 hidden nodes and 10 output nodes. It therefore has the same number of degrees of freedom as the neural net, but is connected differently. The computer is connected by trainable couplings to the 784 pixels of the MNIST digit, whose values are shifted and scaled in order to have zero mean and unit variance (they are also further normalized to ensure that each digit has the same L2 norm). The computer has all-to-all connections between input data and hidden nodes, between input data and output nodes, between hidden nodes and output nodes, and within the set of hidden nodes. In all, it has 63,143 adjustable parameters. The intrinsic energy scale of the computer is set by the couplings $J_2 = J_4 = \kt$; all energies will be displayed in units of $\kt$. 

We constructed idealized teacher trajectories using a noninteracting version of the thermodynamic computer $(J_{ij}^{(0)}=0)$ run at zero temperature $(T^{(0)}=0)$. For a given MNIST digit, the guide biases $b_i^{(0)}$ of the teacher's hidden units were set proportional to the activations $A_i$ of the neural network's hidden nodes, while the guide biases $b_i^{(0)}$ of the output units were set proportional to the shifted activations $2A_i-1$ of the neural network's output nodes. In this way we wish to encourage the output units of the student computer to become positive when they recognize a digit of the relevant class, and negative otherwise. The long-time values of the activations under the teacher trajectories are not directly proportional to the values of the biases, because units' intrinsic responses are nonlinear; in addition, the effective activation function of the thermodynamic computer is different in detail to the hyperbolic tangent function of the neural network\c{whitelam2024thermodynamic}. However, the teacher computer's biases and long-time activations are strongly correlated, and we reason this is sufficient for a classifier, which depends ultimately on the hierarchies of the output activations and not their precise values. Activations were initially zero, $x_i(0)=0$, and the trajectory time was set to be $\tf=1/5$, in units of $\mu^{-1}$, the fundamental clock time of the computer. This time is short enough that the trained computer is out of equilibrium when observed, but long enough for unit activations $x_i(\tf)$ to become nonlinear functions of their inputs\c{whitelam2024thermodynamic}, so providing the expressiveness of a universal approximator.

The student computer was trained using multiple passes through a class-balanced version of the MNIST training set, slightly fewer than $10^6$ trajectories in all. Testing is done by taking the trained computer, showing it an MNIST test-set digit, identifying which output unit is most positive, and comparing that prediction with the digit's class. The computer achieves a top-1 (respectively top-2) test-set accuracy of 92.0\% (respectively 96.7\%) on the standard MNIST test set, when averaged over 10 independent trajectories. For single trajectories, these numbers are respectively 91.7\% and 96.5\%, showing the computer's calculation to be nearly deterministic, despite the stochastic nature of its dynamics. Although there exists a performance gap between the thermodynamic computer and the neural network, this result demonstrates the viability of gradient-descent training for thermodynamic computing. We anticipate that the accuracy of the thermodynamic computer can be improved by tweaking its architecture and by adopting or adapting the enhancements to gradient descent used in deep learning\c{lecun2015deep,schmidhuber2015deep}.

\f{fig2}(a) shows the loss function, the negative log probability with which the student computer would generate the idealized teacher trajectory, as a function of the number of training trajectories $n$. We generate one training trajectory for each digit observed, and pass repeatedly through the training set, updating the student computer's parameters according to Eqs.~\eq{grad1} and \eq{grad2} after each digit (called {\em online learning}). We also show the accuracy $A$ of the partially trained student computer on a 1000-digit subset of the test set, as a function of $n$. This information is not used during training.

Once the student computer is trained, we can observe its behavior when shown various MNIST digits. In \f{fig2}(b) we show (solid lines) three of the trained thermodynamic computer's activations $x_i$ as a function of time, when shown a particular training-set digit. The teacher trajectories are shown as dotted lines. The coincidence of the dotted and solid lines shows the computer to have learned to reproduce the teacher trajectories with reasonable accuracy. Panel (c) compares all the activations of the trained computer, at time $\tf$, with those of the neural network, for two training-set digits. The computer's outputs are averaged over 5 independent trajectories. Blue and red pixels indicate positive and negative values, respectively. While not identical, the correlation between the thermodynamic computer's activations and those of the trained neural network is clear.
 
In \f{fig2}(d) we show how inference works during the operation of the trained thermodynamic computer. When shown the indicated digit, the three outputs of the computer adopt values showing that it predicts the digit, correctly, to be a 0. Panel (e) shows an example in which the computer misclassifies the digit (a 4) as a 0. Its output shows its decision to be split between those two classes. The computer is visibly out of equilibrium, its one-time quantities still evolving. The training process is designed to produce an answer at a specified time, whether or not the computer has attained equilibrium at that time.
 
Finally, \f{fig2}(f) compares the neural-network weights $w_{ij}$ with the thermodynamic computer couplings $J_{ij}$ connecting the 784 input pixels to three representative hidden units. The patterns show clear similarities, with one notable difference: the couplings of the thermodynamic computer break symmetry and take both positive and negative values, reflecting the fact that its input pixels are centered symmetrically about zero.

\section{Energetic considerations}
\begin{figure} 
   \centering
 \includegraphics[width=0.9\linewidth]{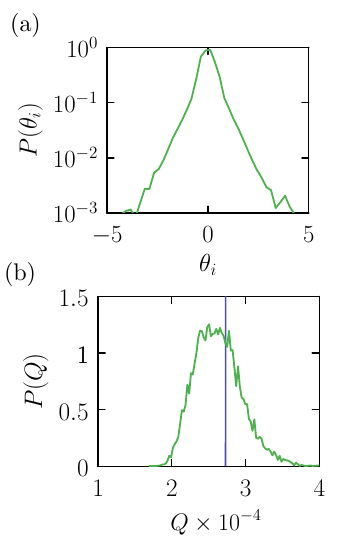} 
   \caption{Histograms of (a) the couplings of the trained thermodynamic computer and (b) the heat it emits while classifying an MNIST digit. In (b), the green distribution is taken over the 10,000-member test set, while the blue distribution is taken over 10,000 trajectories using a single test-set digit. Both horizontal axes are given in units of $\kt$. The overall energy scale of a hardware version of the thermodynamic computer performing MNIST inference would be about $10^5 \kt$. By contrast, a {\em single} multiply-accumulate operation in a neural network using digital hardware costs about $2 \times 10^8 \, \kt$, and MNIST inference using the teacher neural network costs about $5 \times 10^{12}\,\kt$.}
   \label{fig3}
\end{figure}

Being able to train a thermodynamic computer to mimic a neural network enables low-energy machine learning in thermodynamic hardware. To quantify the {\em thermodynamic advantage}
\beq
\label{adv}
{\cal A}_{\rm th}\equiv \frac{E_{\rm digital}}{E_{\rm thermo}},
\eeq
the ratio of energy costs of doing the computation digitally and thermodynamically, we can compare the energy scales of MNIST inference on digital and thermodynamic hardware. 

The basic energy scale of a neural network is the cost of a multiply-accumulate (MAC) operation, in which two numbers are multiplied and added to a register. The energy cost of a MAC varies with hardware, but an order-of-magnitude estimate for a digital implementation is 1 pJ\c{horowitz20141}. Given that $\kt$ at 300 K is roughly $4 \times 10^{-21}\,{\rm J} = 4 \times 10^{-9}\,{\rm pJ}$, each MAC costs about $2 \times 10^8 \, \kt$. The teacher neural network has a 784--32--32--10 architecture, which involves 26,432 MACs. The energy cost of evaluating the neural network, for one MNIST digit, is therefore about $5 \times 10^{12}\,\kt$.

If realized in hardware, the energy cost of the thermodynamic computer would be many orders of magnitude smaller. We can estimate this cost by considering the energy scales of different elements of the computer.

First, the inputs. The input to the thermodynamic computer consists of 784 pixels, each represented by a number of order unity on the scale of $\kt$. In total, therefore, the energy scale of the input is of order $10^3 \,\kt$. If each pixel value must be physically replicated across its downstream connections -- which may be necessary for some hardware designs but not others -- then this cost would increase to $10^4$--$10^5 \,\kt$. 

Second, the couplings. In \f{fig3}(a) we show a histogram $P(\theta_i)$ of the values of the $63,143$ parameters $\theta_i$ of the trained thermodynamic computer. The horizontal axis is given in units of $\kt$. The distribution is far from Gaussian, with a sharp peak around zero and broad shoulders, reflecting the heterogeneous interactions required to reproduce the neural network's activations. The typical scale of the couplings is about $\kt$. In the worst-case scenario, if all couplings have to be supplied via external work during inference, then the associated energy budget would be of order $10^5 \, \kt$. However, in existing hardware realizations of thermodynamic computers the couplings are fixed physical parameters, and no external work is required to supply them during inference\c{melanson2025thermodynamic,aifer2025solving}. In this case the energy budget for the couplings is zero.

Third, we can measure the heat $Q$ dissipated to the thermal bath during inference, i.e. during a trajectory of the trained computer when presented with an MNIST digit. The heat emitted during a trajectory $\{\x(t_k)\}_{k=0}^{K}$ is
\beq
Q = \sum_{k=0}^{K-1} \Big[ V_{\tt}(\x(t_k)) - V_{\tt}(\x(t_{k+1})) \Big],
\eeq
where $V_{\tt}(\x)$ is given by \eqq{pot}. In \f{fig3}(b) we show a histogram $P(Q)$ of the heat emitted while classifying an MNIST digit, taken over all 10,000 digits of the MNIST test set (green). The horizontal axis is given in units of $\kt$. The heat emitted during inference is of order $10^4 \, \kt$. We also show (blue) the histogram of heat values for 10,000 trajectories using a single MNIST digit. This distribution is much narrower, and on the scale of the figure appears as a single peak. The comparison demonstrates that the width of the green distribution is set primarily by the differing energetic costs of inference for different input digits, rather than by stochastic fluctuations between trajectories of the same digit.

These three considerations suggest that a hardware implementation of our thermodynamic computer could perform inference with an energy cost of order $10^5 \kt$, giving a thermodynamic advantage \eq{adv} of more than $10^7$.

  \section{Conclusions}

We have shown that gradient descent can be used to train a thermodynamic computer to perform computations on a clock, at fixed observation times and in general out of equilibrium. In a recent paper we trained a model thermodynamic computer using gradient descent to implement a generative model\c{whitelam2025generative}. Here we have extended this approach, showing that gradient descent can also train a thermodynamic computer to mimic the behavior of a target neural network trained to perform an arbitrary calculation. The training scheme follows a teacher-student framework: a neural network provides reference activations, which are approximated by idealized Langevin trajectories. By differentiating the Onsager-Machlup functional, we maximize the likelihood that the thermodynamic computer generates these idealized trajectories, thereby shaping its dynamics to reproduce the target computation. We demonstrated this method on an image-classification task.

We had previously shown that a genetic algorithm can be used to train a thermodynamic computer to perform machine-learning tasks\c{whitelam2024thermodynamic}. Genetic algorithms are as effective as gradient descent, able to reach similar values of loss, but are slower and more costly computationally: for the 10-class MNIST problem, the genetic algorithm required the generation of about $10^{13}$ trajectories; for GD, the number is fewer than $10^6$. One advantage of GAs, however, is that they can be applied directly to a device in hardware\c{sabattini2025towards}. One could therefore combine these methods, training a digital model of a thermodynamic computer using gradient descent, and, if necessary, retraining its hardware realization using a GA. The latter step is potentially useful if the device production process results in a stochastic distribution of component characteristics, sometimes called {\em device noise}.

We also estimated the energy costs associated with the teacher neural network and a hardware implementation of the trained thermodynamic computer, and concluded that the thermodynamic advantage -- the ratio of the digital and thermodynamic energy costs -- exceeds seven orders of magnitude. This simple example illustrates the considerable potential of thermodynamic computing. The neural network is slightly more accurate than the thermodynamic computer, but this gap in performance can almost certainly be narrowed by fine-tuning the computer's training procedure. For example, it may be possible to construct teacher trajectories that are easier for a student computer to mimic, and the gradient-descent method could be augmented with techniques commonly used in neural-network training, including momentum and weight decay.

This paper demonstrates the feasibility of applying machine-learning methodology to thermodynamic devices. It establishes gradient descent as a practical training method for thermodynamic computing, opening the door to training energy-efficient thermodynamic hardware to perform the broad range of tasks addressed by modern machine learning.

{\em Acknowledgments---} This work was done at the Molecular Foundry, supported by the Office of Science, Office of Basic Energy Sciences, of the U.S. Department of Energy under Contract No. DE-AC02-05CH11231, and partly supported by US DOE Office of Science Scientific User Facilities AI/ML project ``A digital twin for spatiotemporally resolved experiments''.


%

\end{document}